\documentclass[%
 reprint,
superscriptaddress,
 amsmath,amssymb,
 aps,
pra,
]{revtex4-2}
\usepackage[utf8]{inputenc}
\usepackage{physics}
\usepackage{graphicx}
\usepackage{xcolor}
\usepackage{float}
\newcommand{\coh}[1]{\expandafter\hat\sigma_{#1}}
\newcommand{\conj}[1]{{#1}^{\dagger}}
\newcommand{\comj}[1]{{#1}^{*}}
\usepackage{csquotes}
 
\begin{document}
\title{Observation of cross phase modulation in cold atom gradient echo memory}

\author{Anthony C. Leung}
    \affiliation{ARC Centre of Excellence for Quantum Computation and Communication Technology}
    \affiliation{Department of Quantum Science and Technology, Research School of Physics, The Australian National University, Acton, ACT 2601, Australia}
\author{K. S. Ida Melody}
    \affiliation{ARC Centre of Excellence for Quantum Computation and Communication Technology}
    \affiliation{Department of Quantum Science and Technology, Research School of Physics, The Australian National University, Acton, ACT 2601, Australia}
\author{Aaron D. Tranter}
    \affiliation{ARC Centre of Excellence for Quantum Computation and Communication Technology}
    \affiliation{Department of Quantum Science and Technology, Research School of Physics, The Australian National University, Acton, ACT 2601, Australia}
\author{Karun V. Paul}
    \affiliation{ARC Centre of Excellence for Quantum Computation and Communication Technology}
    \affiliation{Department of Quantum Science and Technology, Research School of Physics, The Australian National University, Acton, ACT 2601, Australia}
\author{Geoff T. Campbell}
    \affiliation{ARC Centre of Excellence for Quantum Computation and Communication Technology}
    \affiliation{Department of Quantum Science and Technology, Research School of Physics, The Australian National University, Acton, ACT 2601, Australia}
\author{Ping Koy Lam}
    \affiliation{ARC Centre of Excellence for Quantum Computation and Communication Technology}
    \affiliation{Department of Quantum Science and Technology, Research School of Physics, The Australian National University, Acton, ACT 2601, Australia}
\author{Ben C. Buchler}
    \email{Corresponding author; ben.buchler@anu.edu.au}
    \affiliation{ARC Centre of Excellence for Quantum Computation and Communication Technology}
    \affiliation{Department of Quantum Science and Technology, Research School of Physics, The Australian National University, Acton, ACT 2601, Australia}


\date{\today}

\begin{abstract}
Strong nonlinear interactions between single photons have important applications in optical quantum information processing.  Demonstrations of these interactions in cold atomic ensembles have largely been limited to exploiting slow light generated using electromagnetically induced transparency (EIT).  However, these EIT implementations have limited achievable phase shifts due to spontaneous emission.  Here, we demonstrate and characterize a scheme free from these limitations using gradient echo memory with inferred single photon phase shifts of $0.07\pm0.02$ $\mu$rad.  Excellent agreement with theoretical modelling was observed.  Degradation of memory efficiency was observed for large phase shifts but strategies to overcome that are presented.
\end{abstract}

\maketitle
\section{Introduction}
Facilitating strong interactions between photons is an important step towards realizing optical quantum computing \cite{Milburn_1989}.  This can be achieved via materials possessing a very strong nonlinearity causing a photon to impart a phase shift onto another through cross phase modulation (XPM).  This process can be used for many interesting applications such as universal quantum gates \cite{Milburn_1989}, quantum non-demolition measurements of photon numbers \cite{Imoto_Haus_Yamamoto_1985} and cluster state generation \cite{Kim_Lee_Ji_Nha_Anisimov_Dowling_2015}.

Various proposals have demonstrated examples of strong nonlinear interactions using different platforms.  Early promising systems took advantage of cavity quantum electrodynamics (cQED) using single atoms \cite{Turchette_Hood_Lange_Mabuchi_Kimble_1995} and quantum dots \cite{Fushman_Englund_Faraon_Stoltz_Petroff_Vuckovic_2008}.  Over the last few years, successful demonstrations of large phase shifts were observed using cQED applied to an atomic ensemble to get phase shifts of up to $\pi/3$ per postselected single photon \cite{Beck_Hosseini_Duan_Vuletic_2016}, Rydberg  blockades \cite{Tiarks_Schmidt_Rempe_Durr_2016} for $\pi$ per photon and hollow core fibres filled with atomic gas \cite{Venkataraman_Saha_Gaeta_2013} for $0.3$mrad per photon. 

The main approach for generating XPM in atomic ensemble systems has been through optical dispersion engineering \cite{Chen_Wang_Wang_Yu_2006, Lo_Su_Chen_2010,Lo_Chen_Su_Chen_Chen_Chen_Yu_Chen_2011}.  The slow-light effect in EIT can be used to strengthen the nonlinear interaction in the medium as a.c.-Stark shift imparts a phase shift and creates XPM.  Phase shifts of up to 18 $\mu$rad per post-selected single photon have been observed \cite{Feizpour_Hallaji_Dmochowski_Steinberg_2015}.  It has been shown, however, that the magnitude of phase shift obtainable between two travelling photons via EIT is limited \cite{Gea-Banacloche_2010}.  This limit could only be overcome by operating the medium as a quantum memory to avoid simultaneously propagating optical fields or in the coherent photon conversion regime (a four-wave mixing effect that causes atomic population oscillations) which was recently demonstrated \cite{sagona-stophelConditionalPPhaseShift2020}.

In this paper, we measured the nonlinearity achieved by storing one light field using gradient echo memory (GEM) \cite{hosseiniHighEfficiencyCoherent2011, choHighlyEfficientOptical2016} and propagating the other field through the medium during the storage phase.  We infer 0.07 $\mu$rad of phase shift per single photon.  While this phase shift is fairly modest, we present prospects for increasing the size of the phase shift through changes to experimental design.

\section{Method}
\subsection{Preparation of ensemble}
A cloud of \textsuperscript{87}Rb atoms is captured and cooled using a magneto-optical trap (MOT).  A 2D quadrupole field is used to create an elongated geometry to maximize optical depth along the propagation path of the probe (optical depth of 450 on probe transition of $D_1$ $\ket{5S_{1/2},F{=}2} \rightarrow \ket{5P_{1/2}, F'{=}1}$).  The optical table is surrounded by three pairs of orthogonal Helmholtz coils to cancel out static background fields.  The experimental sequence was synchronized to the 50Hz mains power to ensure environmental consistency from run to run.   Details on the ensemble preparation can be found in \cite{choHighlyEfficientOptical2016, tranter2018}.

\subsection{Measuring XPM using GEM}
\begin{figure*}
\centering
\includegraphics[scale=0.8]{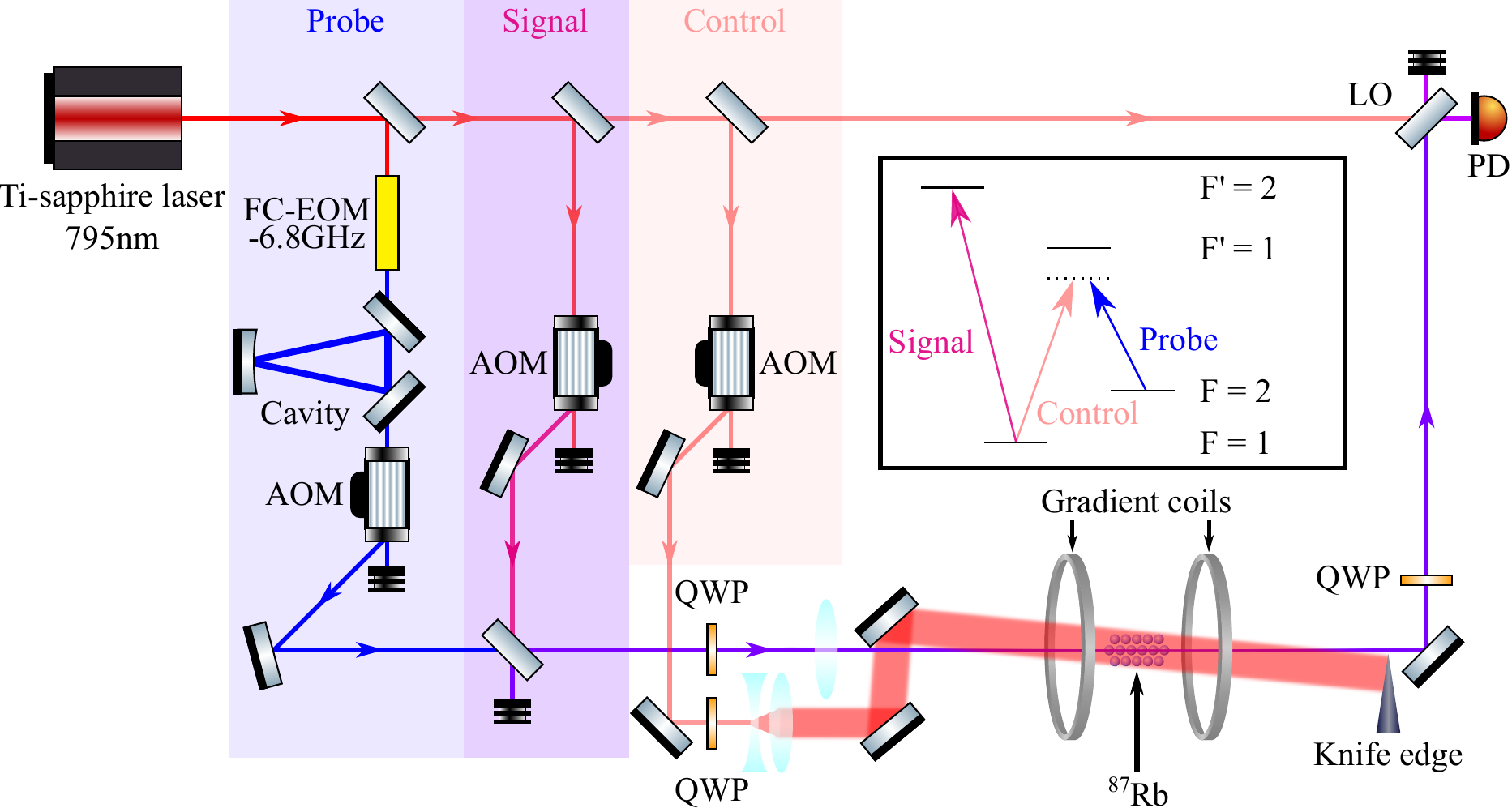}
\caption{Experimental setup.  A Ti:Sapphire laser generated both the control and signal fields using acousto-optic modulators (AOM).  The probe field separated at 6.8 GHz was generated through modulation using a fibre coupled electro-optic modulator (FC-EOM) with a cavity to select the frequency shifted sideband.  The probe and signal were combined on a beamsplitter and sent through a quarter wave plate (QWP) to become circularly polarized before the probe is stored in a cold atomic ensemble of $^{87}$Rb atoms.  Gradient coils control the magnetic field gradient across the ensemble to store and recall the probe from the $\ket{5S_{1/2},F{=}1}$ state via the control beam.  The recalled probe was combined with a local oscillator (LO) beam for phase measurements via heterodyne detection on a photodetector (PD).  The inset shows the three level GEM memory scheme formed by the probe and control fields along with the signal field.}
\label{fig:setup}
\end{figure*}

The experimental setup is shown in figure \ref{fig:setup}.  A weak probe was focused into the MOT along the elongated axis with a strong control field intersecting at an angle of $0.45\pm0.05^{\circ}$.  Three level GEM was implemented using these beams as illustrated in the inset of figure \ref{fig:setup}.  The probe and control fields were red detuned by $\Delta = 208$ MHz from the $\ket{5S_{1/2},F{=}2} \rightarrow \ket{5P_{1/2}, F'{=}1}$ and $\ket{5S_{1/2},F{=}1} \rightarrow \ket{5P_{1/2}, F'{=}1}$ of the \textsuperscript{87}Rb $D_1$ line respectively.  A pair of coils generated the magnetic field gradient required to inhomogeneously broaden the atomic transition to 200 kHz in the MOT to operate GEM. 

\begin{figure*}
\centering
\includegraphics[scale=0.65]{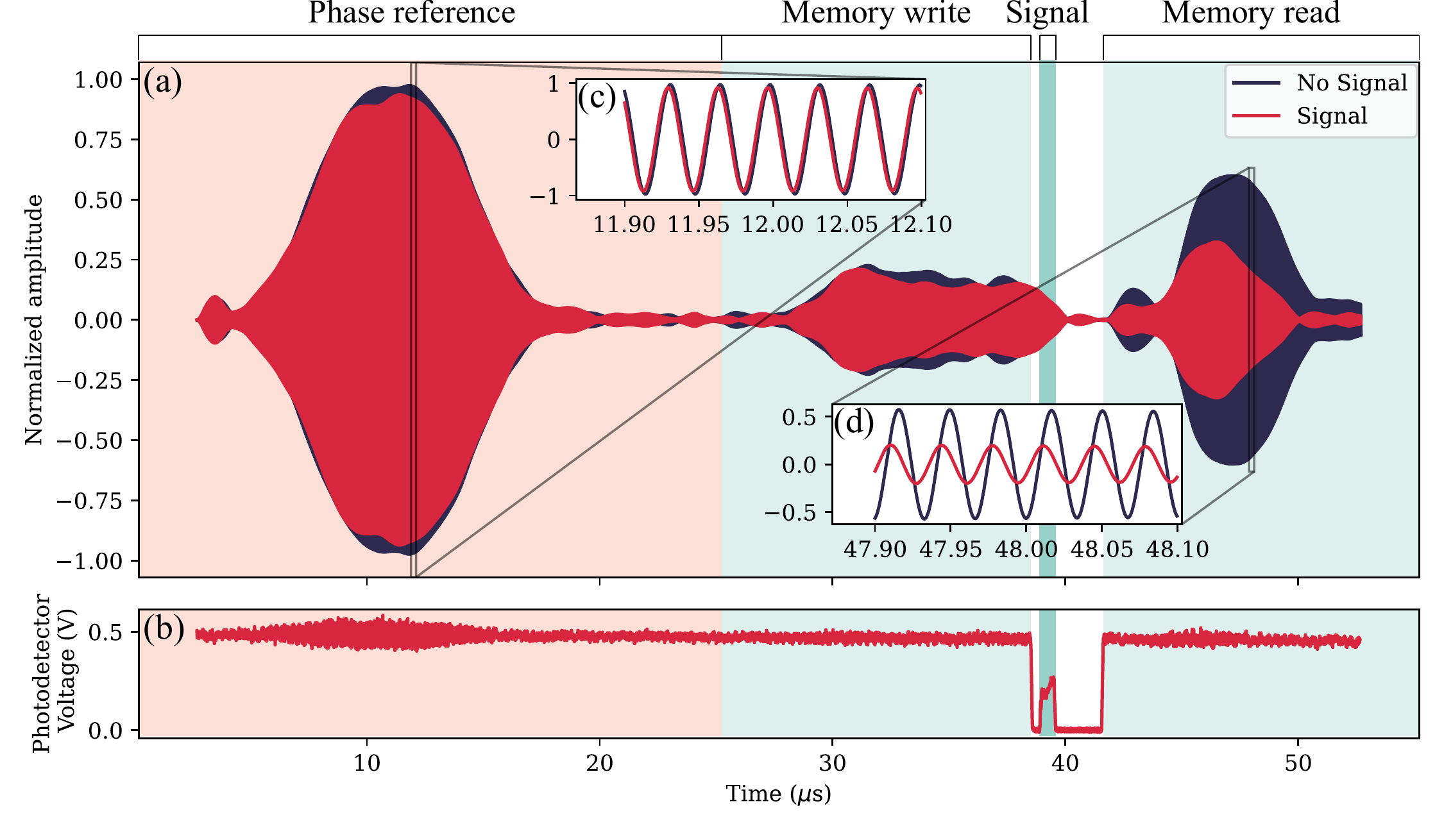}
\caption{Heterodyne measurement data for 3.7 pJ of signal energy at a detuning of $\delta = -8.7$ MHz.  a) Normalized data that has been filtered down to only the heterodyne beat frequency.  A bright pulse during the phase reference section is used to account for phase drifts between the data sets with and without signal.  A probe pulse is stored and recalled during \textquote{Memory write} and \textquote{Memory read} between which a signal pulse is sent through the memory. b) Raw data from the photodetector; the local oscillator was switched off during storage to measure the signal pulse.  The phase shift due to the signal pulse is shown with c) and d).  c) shows the phase reference pulse being in phase for both traces with and without the signal pulse and d) shows the phase of the recall being shifted due to the signal pulse.}
\label{fig:pulse}
\end{figure*}

The phase of the recall was measured using heterodyne detection.  An illustration of the pulse sequence used to perform this measurement is shown in figure \ref{fig:pulse}.  Before storage, a probe pulse was sent through the MOT with the control field switched off. Due to the detuning of the probe, this light passed through the atomic ensemble without interacting and was used as a phase reference.  After switching the control field on, a probe pulse was stored for 13 $\mu$s and recalled by switching the current in the gradient coils.  The XPM was generated by the signal pulse of duration $\tau$ and addressed the  $F{=}1 \rightarrow F'{=}2$ \textsuperscript{87}Rb $D_1$ transition with a detuning $\delta$.  This field was mode-matched to the probe and both coupled into the same fibre before being sent to the MOT to maximise its impact. The amount of XPM was measured by comparing recall from the memory with and without the signal pulse.


The a.c.-Stark shift can be integrated over the signal pulse duration for a simple analytical expression of the expected cross phase modulation \cite{Chen_Wang_Wang_Yu_2006}:

\begin{equation}
    \phi_{\textnormal{XPM}} = \frac{\Omega_s^2\delta\tau}{2\left(\gamma^2+\delta^2\right)}
    \label{eq:xpm}
\end{equation}
\noindent where $\Omega_s$, $\tau$ and $\gamma$ are the signal Rabi frequency, signal pulse duration and transition linewidth respectively.

\section{Cross phase modulation measurement}
\begin{figure}[h]
\centering
\includegraphics[scale=0.8]{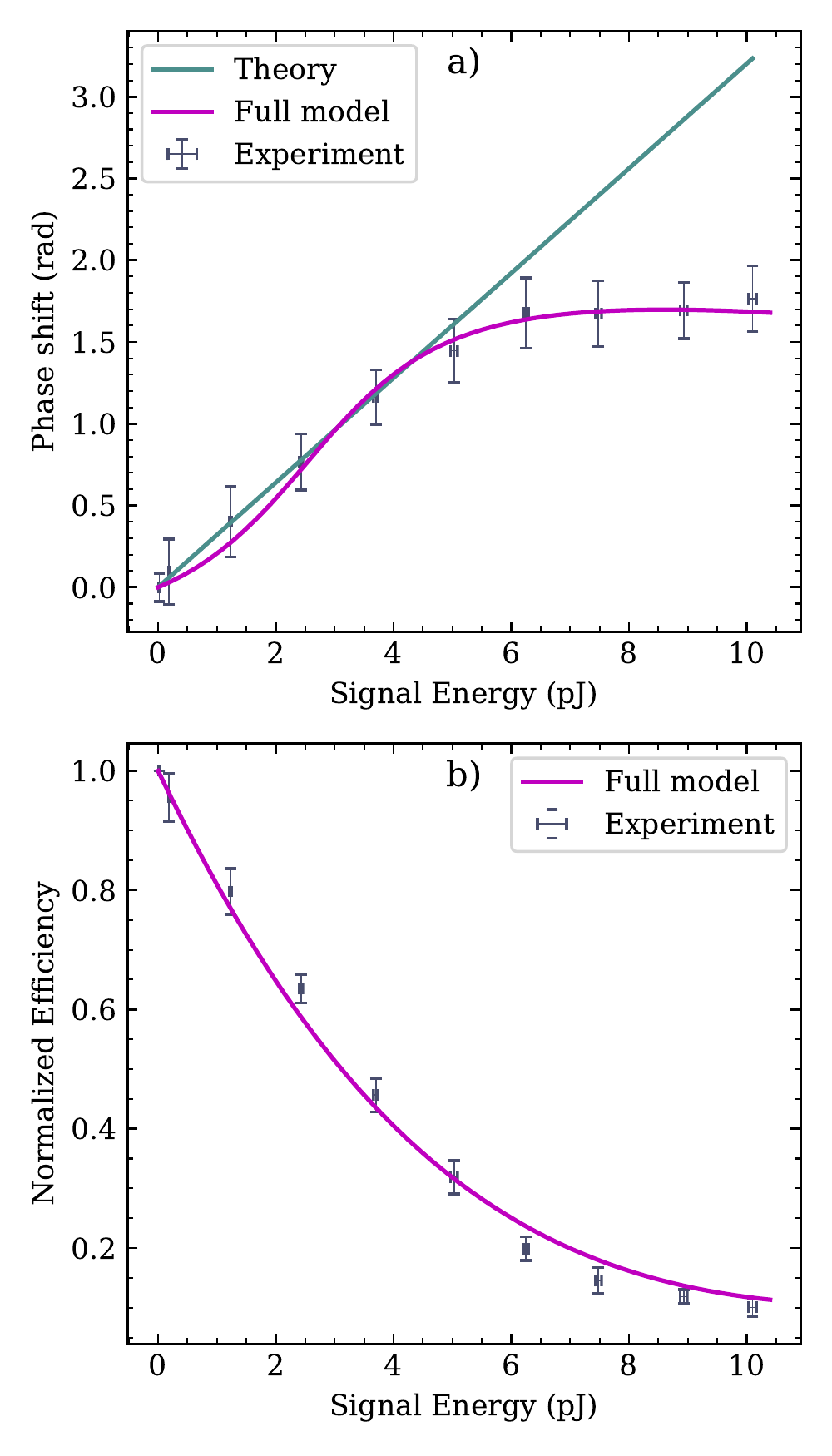}
\caption{a) Measured phase shift due to cross phase modulation at a series of signal pulse energies along with the theoretically predicted phase shifts using (1) at $\delta = -8.7$ MHz and the phase shift expected using a full model that accounts for an impure initial state using Maxwell-Bloch equations. b) Normalized recall efficiency for increasing signal pulse energies overlaid with the efficiency expected by the full model Maxwell-Bloch equations.}
\label{fig:evp}
\end{figure}

Figure \ref{fig:evp}a shows the measured phase shift for a series of increasing signal pulse energies, along with a plot of what is expected using (1), at $\delta = -8.7$ MHz.  The energy of the pulse was varied by changing the pulse duration and measured by integrating over the photodetector signal over that duration.  The beam size in this experiment was measured to be 150$\pm$40 $\mu$m.  Good agreement was observed between the data and the model up to about 5 pJ signal energy (when fitted with a beam size of 190 $\mu$m which is within error).  Above this energy, there is significant discrepancy between the simple model and experimental data.  It appears as though the achievable phase shift saturates in a way that is not captured by (1).

To fully understand this phase saturation effect, the full model Maxwell-Bloch equations of this four level system were numerically solved.  These equations and more details on how they were used can be found in the Appendix (A.1).  It was observed that this saturation effect only occurs when there is some impurity in the initial state of the system; the storage state ($\ket{5S_{1/2},F{=}1}$) had some population before the memory operation.  Due to the relatively small control detuning of this experiment, the control field was capable of incoherently scattering a very small level of light into the probe mode.  As the signal energy is increased, the size of the recall is reduced as seen in figure \ref{fig:evp}b but the level of light incoherently scattered by the control field remains constant.  The phase saturation occurs when the control scattered light started to dominate over the reduced recall signal.  The initial population in the storage state was calculated by measuring the signal absorption at various signal frequencies (figure \ref{fig:sigabs} in the appendix) and was found to be 2\% of the total population (the simulation ignored other $m_F$ states in the same manifold so the 2\% population could be interpreted as there being 2 atoms in the storage state for every 98 atoms in the initial ground state).  As plotted in the figure \ref{fig:evp}a, the full numerical model using the above initial atomic population state ratio predicts phase shift saturation at $\sim$1.5 rad which agrees well with the experimental data.  The precise parameters of this model were optimized to fit experimental results using the same machine learner as in \cite{tranter2018} within realistic bounds set by the experiment.

\begin{figure}[h]
\centering
\includegraphics[scale=0.7]{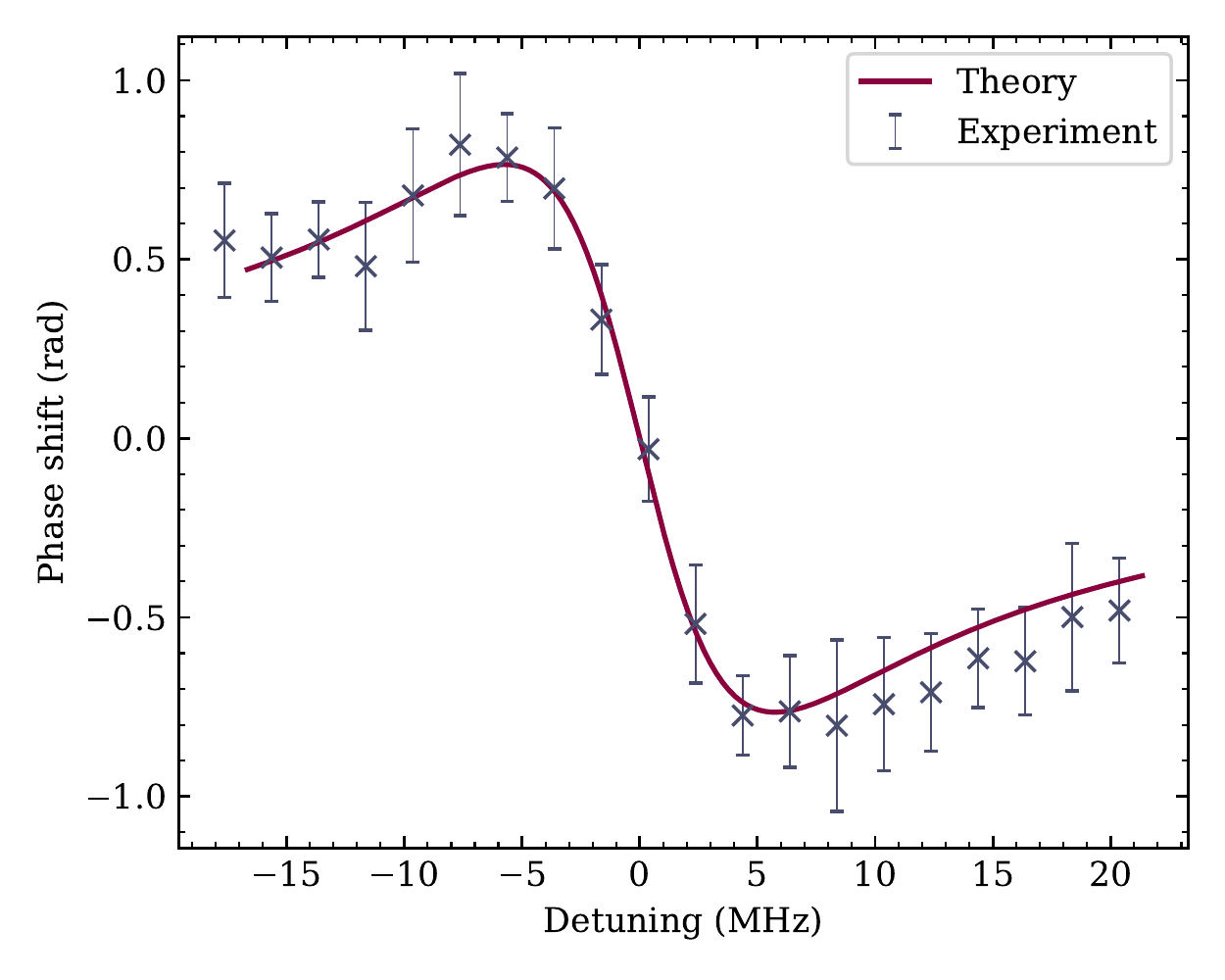}
\caption{XPM phase shifts for different detunings, $\delta$, from the Stark shift transition at 2.3 pJ of signal energy overlaid with the theoretically predicted phase shifts using (1). }
\label{fig:fvp}
\end{figure}

The effect of changing the signal detuning was also investigated at 2.3 pJ signal energies as shown in figure \ref{fig:fvp} and good agreement between theory and experiment was also observed.  The error in the phase shift in both figures \ref{fig:evp}a and \ref{fig:fvp} were attributed to phase fluctuations at the heterodyne detector; the probe and local oscillator had to travel through separate optical fibres so their relative phase suffered from significant fluctuations, which can reach more than $2\pi$ rad from data set to data set.  The phase shift uncertainty in each data point on both plots were propagated from the standard deviation over 20 data-sets of the phase difference between the reference pulse and the recall (for with and without signal pulse).  

\section{Discussion}
Extrapolating from figure \ref{fig:evp}a, a signal pulse containing a single photon will cause 0.07$\pm$0.02 $\mu$rad of phase shift in the stored probe.  This level of phase shift is still not large enough to perform the weak Kerr nonlinearities gate protocol Munro et al. proposed in \cite{Munro_Nemoto_Spiller_2005} that requires $10^{-5}$ rad.  Moreover our experiment shows absorption of the probe and signal fields that would further disadvantage a quantum gate operation.  We will now consider how these limitations may be addressed.

As seen in figure \ref{fig:evp}b, memory efficiency drops as the signal energy increases.  This is  because the atoms in the storage state ($F{=}1$) are pumped by the signal field leading to incoherent absorption of both probe and signal fields.  This effect was also observed in the numerical simulations using the Maxwell-Bloch equations.  The effect of the signal energy on the memory efficiency was particularly sensitive to memory bandwidth.  This was because the signal beam was focused through the ensemble and therefore had a spatially varying intensity across the ensemble; this imparted a spatially varying phase shift.  This hindered the rephasing of the atoms as the stored state was recalled.  Due to the nature of GEM storing frequency components of the input pulse across a spatial Fourier transform, smaller memory bandwidths meant the written atomic coherence was more spread out through the ensemble.  This caused the signal pulse to degrade the memory efficiency more for smaller memory bandwidths.

This probe and signal absorption, however, is not an inherent property of the a.c.-Stark shift scheme presented here.  The signal pulse reducing memory efficiency could be mitigated by either using larger memory bandwidths to localize the stored pulse in a smaller physical volume around the signal beam focus or use a less tightly focused signal beam.  As mentioned previously, it was determined that the atomic ensemble was initialized with some residual population in the $F{=}1$ storage state of the memory.  On the basis of our numerical model, this state impurity is responsible for most of the signal absorption (figure \ref{fig:sigabs}), as well as the saturation of the phase shift (seen in figure~\ref{fig:evp}a).  The signal absorption could, therefore, be strongly mitigated if the ensemble were prepared purely in the $F{=}2$ state.  An effectively 100\% pure initial state could be obtained via RF evaporation which is commonly used to create Bose-Einstein condensates \cite{andersonObservationBoseEinsteinCondensation1995}.  Crucially, the atoms transferred to the storage state via the memory operation will have a negligible effect on the signal absorption in a quantum gate application where the probe is a weak excitation. Signal absorption can be further suppressed if larger signal detunings were used at the expense of some phase shift magnitude. 


Since there are clear ways to mitigate signal absorption, absorption of the probe light will be the limit on fidelity when using this system as a quantum gate. State impurity in this case will not be a factor since the scheme uses Fock states with very few photons in the signal field, as seen in figure~\ref{fig:evp}b.  The GEM protocol has to date shown a best efficiency of 87\% \cite{choHighlyEfficientOptical2016}.  This will be a limit if we are seeking to build unconditional quantum gates, where the loss of a photon means a loss of fidelity. In principle, GEM can attain efficiencies approaching 100\%.  The current operational limit, we believe, is due to inhomogeneity in the control field.  In future work we will address this issue as we aim for higher GEM efficiency.  Machine learning \cite{tranter2018} will also be explored as way to optimise the pulse timing and shaping.

In terms of the achievable nonlinearity, a simple path forward towards larger phase shift is to focus the signal field more tightly at the MOT to achieve higher signal intensities.  The relatively large beam size in this experiment leaves much room for improvement.  If the beam waist was tightened to $1 \mu$m, up to $2.5$ mrad of single photon phase shift could be observed.  Further improvements may also be observed if slow light or stationary light effects are applied to the signal field.

\section{Conclusion}
In conclusion, we have characterized the phase shift achievable via cross phase modulation facilitated by a cold-atom quantum memory.  0.07$\pm$0.02 $\mu$rad of phase shift per single photon was inferred by our measurement with excellent agreement with our modeling.  This measurement is a promising first step towards using quantum memories as deterministic quantum gates which is free from the limitations detailed in \cite{Gea-Banacloche_2010} for travelling wave cross phase modulation implementations.  Further improvements in phase shift of orders of magnitude is possible with this scheme through tighter beam focusing and application of slow or stationary light.

This research was conducted by the Australian Research Council Centres of Excellence Centre for Quantum Computation and Communication Technology (Grant No. CE170100012).
\bibliography{bibliography.bib}

\providecommand{\newblock}{}
\begin{thebibliography}{10}
\expandafter\ifx\csname url\endcsname\relax
  \def\url#1{{\tt #1}}\fi
\expandafter\ifx\csname urlprefix\endcsname\relax\def\urlprefix{URL }\fi
\providecommand{\eprint}[2][]{\url{#2}}

\bibitem{Milburn_1989}
Milburn G~J 1989 {\em Physical Review Letters\/} {\bf 62} 2124–2127 ISSN
  0031-9007

\bibitem{Imoto_Haus_Yamamoto_1985}
Imoto N, Haus H~A and Yamamoto Y 1985 {\em Physical Review A\/} {\bf 32}
  2287–2292 ISSN 0556-2791

\bibitem{Kim_Lee_Ji_Nha_Anisimov_Dowling_2015}
Kim J, Lee J, Ji S~W, Nha H, Anisimov P~M and Dowling J~P 2015 {\em Optics
  Communications\/} {\bf 337} 79–82 ISSN 00304018

\bibitem{Turchette_Hood_Lange_Mabuchi_Kimble_1995}
Turchette Q~A, Hood C~J, Lange W, Mabuchi H and Kimble H~J 1995 {\em Physical
  Review Letters\/} {\bf 75} 4710–4713 ISSN 0031-9007, 1079-7114

\bibitem{Fushman_Englund_Faraon_Stoltz_Petroff_Vuckovic_2008}
Fushman I, Englund D, Faraon A, Stoltz N, Petroff P and Vuckovic J 2008 {\em
  Science\/} {\bf 320} 769–772 ISSN 0036-8075, 1095-9203

\bibitem{Beck_Hosseini_Duan_Vuletic_2016}
Beck K~M, Hosseini M, Duan Y and Vuletić V 2016 {\em Proceedings of the
  National Academy of Sciences\/} {\bf 113} 9740–9744 ISSN 0027-8424,
  1091-6490

\bibitem{Tiarks_Schmidt_Rempe_Durr_2016}
Tiarks D, Schmidt S, Rempe G and Dürr S 2016 {\em Science Advances\/} {\bf 2}
  e1600036 ISSN 2375-2548

\bibitem{Venkataraman_Saha_Gaeta_2013}
Venkataraman V, Saha K and Gaeta A~L 2013 {\em Nature Photonics\/} {\bf 7}
  138–141 ISSN 1749-4885, 1749-4893

\bibitem{Chen_Wang_Wang_Yu_2006}
Chen Y~F, Wang C~Y, Wang S~H and Yu I~A 2006 {\em Physical Review Letters\/}
  {\bf 96} 043603 ISSN 0031-9007, 1079-7114

\bibitem{Lo_Su_Chen_2010}
Lo H~Y, Su P~C and Chen Y~F 2010 {\em Physical Review A\/} {\bf 81} 053829 ISSN
  1050-2947, 1094-1622

\bibitem{Lo_Chen_Su_Chen_Chen_Chen_Yu_Chen_2011}
Lo H~Y, Chen Y~C, Su P~C, Chen H~C, Chen J~X, Chen Y~C, Yu I~A and Chen Y~F
  2011 {\em Physical Review A\/} {\bf 83} 041804 ISSN 1050-2947, 1094-1622

\bibitem{Feizpour_Hallaji_Dmochowski_Steinberg_2015}
Feizpour A, Hallaji M, Dmochowski G and Steinberg A~M 2015 {\em Nature
  Physics\/} {\bf 11} 905–909 ISSN 1745-2473, 1745-2481

\bibitem{Gea-Banacloche_2010}
Gea-Banacloche J 2010 {\em Physical Review A\/} {\bf 81} 043823 ISSN 1050-2947,
  1094-1622

\bibitem{sagona-stophelConditionalPPhaseShift2020}
{Sagona-Stophel} S, Shahrokhshahi R, Jordaan B, Namazi M and Figueroa E 2020
  {\em Physical Review Letters\/} {\bf 125} 243601

\bibitem{hosseiniHighEfficiencyCoherent2011}
Hosseini M, Sparkes B, Campbell G, Lam P and Buchler B 2011 {\em Nature
  Communications\/} {\bf 2} ISSN 2041-1723
  \urlprefix\url{http://www.nature.com/articles/ncomms1175}

\bibitem{choHighlyEfficientOptical2016}
Cho Y~W, Campbell G~T, Everett J~L, Bernu J, Higginbottom D~B, Cao M~T, Geng J,
  Robins N~P, Lam P~K and Buchler B~C 2016 {\em Optica\/} {\bf 3} 100 ISSN
  2334-2536

\bibitem{tranter2018}
{Tranter} A~D, {Slatyer} H~J, {Hush} M~R, {Leung} A~C, {Everett} J~L, {Paul}
  K~V, {Vernaz-Gris} P, {Lam} P~K, {Buchler} B~C and {Campbell} G~T 2018 {\em
  Nature Communications\/} {\bf 9} 4360

\bibitem{Munro_Nemoto_Spiller_2005}
Munro W~J, Nemoto K and Spiller T~P 2005 {\em New Journal of Physics\/} {\bf 7}
  137–137 ISSN 1367-2630

\bibitem{andersonObservationBoseEinsteinCondensation1995}
Anderson M~H, Ensher J~R, Matthews M~R, Wieman C~E and Cornell E~A 1995 {\em
  Science\/} {\bf 269} 198–201

\end{thebibliography}
\clearpage
\onecolumngrid

\appendix
\section*{Appendix}
\section{Maxwell-Bloch equations}
\begin{figure}[h]
\centering
\includegraphics[scale=1.1]{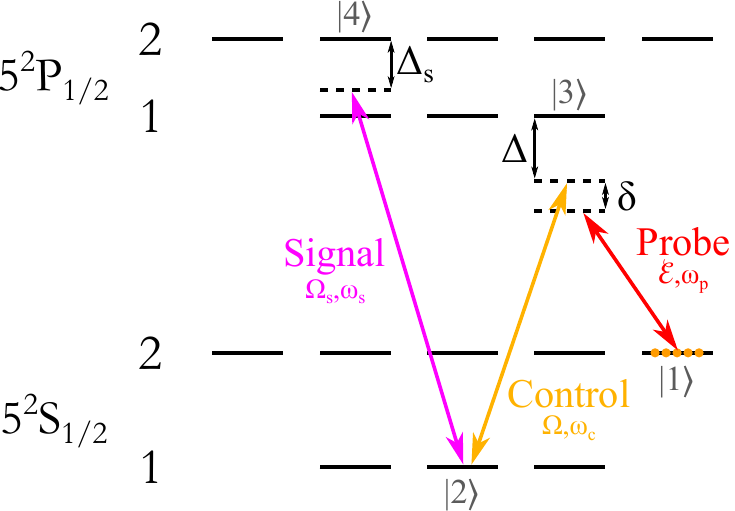}
\caption{The level scheme used to generate the Maxwell-Bloch equations.}
\label{fig:xmds}
\end{figure}

The parameters in the Maxwell-Bloch equations are labelled in figure \ref{fig:xmds}.  These equations have been generated using the atomic and interaction Hamiltonians and simplified using the rotating wave approximation.  Note that in the context of operating GEM, the one photon detuning of the probe, $\delta$, is replaced with the spatially varying gradient term, $\eta z$, caused by the magnetic field gradient inducing a linearly varying Zeeman shift.

\begin{subequations}
\begin{align}
\begin{split}
\partial_t\coh{11} &=-i\Gamma\sqrt{d}(\mathcal{E}\coh{31}-\conj{\mathcal{E}}\coh{13})+\frac{\Gamma}{2}\coh{33}
\end{split}\\
\begin{split}
\partial_t\coh{22} &=i\left(\comj{\Omega}\coh{23}-\Omega\coh{32}+\comj{\Omega_s}\coh{24}-\Omega_s\coh{42}\right)+\frac{\Gamma}{12}\left(\coh{33}+\sqrt{3}(\coh{34}+\coh{43})+3\coh{44}\right)
\end{split}\\
\begin{split}
\partial_t\coh{33}&=i\left(-\comj{\Omega}\coh{23}+\Omega\coh{32}-\Gamma\sqrt{d}(\conj{\mathcal{E}}\coh{13}-\mathcal{E}\coh{31})\right)-\frac{\Gamma}{24}\left(14\coh{33}+\sqrt{3}(\coh{34}+\coh{43})\right)
\end{split}\\
\begin{split}
\partial_t\coh{44}&=-i\left(\comj{\Omega_s}\coh{24}-\Omega_s\coh{42}\right)-\frac{\Gamma}{24}\left(\sqrt{3}(\coh{34}+\coh{43})+6\coh{44}\right)
\end{split}\\
\begin{split}
\partial_t\coh{12}&=-i\left(\comj{\Omega}\coh{23}-\comj{\Omega_s}\coh{14}+\Gamma\sqrt{d}\mathcal{E}\coh{32}+\eta z \coh{12}\right)-\frac{\Gamma}{6\sqrt{2}}\left(\sqrt{3}\coh{33}+3\coh{34}\right)
\end{split}\\
\begin{split}
\partial_t\coh{31}&=i\left(-\comj{\Omega}\coh{21}+\Gamma\sqrt{d}\conj{\mathcal{E}}\left(\coh{33}-\coh{11}\right)+\left(\eta z-\Delta\right)\coh{31}\right) -\frac{\Gamma}{24}\left(7\coh{31}+\sqrt{3}\coh{41}\right)
\end{split}\\
\begin{split}
\partial_t\coh{32}&=i\left(\comj{\Omega}\left(\coh{33}-\coh{22}\right)-\Gamma\sqrt{d}\conj{E}\coh{12}+\comj{\Omega_s}\coh{34}-\Delta\coh{32}\right)-\frac{\Gamma}{24}\left(7\coh{32}+\sqrt{3}\coh{42}\right)
\end{split}\\
\begin{split}
\partial_t\coh{41}&=i\left(\Gamma\sqrt{3}\conj{\mathcal{E}}\coh{43}-\comj{\Omega_s}\coh{21}+\left(\eta z-\Delta_s\right)\coh{41}\right)-\frac{\Gamma}{24}\left(\sqrt{3}\coh{31}+3\coh{41}\right)
\end{split}\\
\begin{split}
\partial_t\coh{42}&=i\left(\comj{\Omega}\coh{43}+\comj{\Omega_s}\left(\coh{44}-\coh{22}\right)-\Delta\coh{42}\right)-\frac{\Gamma}{24}\left(\sqrt{3}\coh{32}+3\coh{42}\right)
\end{split}\\
\begin{split}
\partial_t\coh{43}&=i\left(\Omega\coh{42}-\comj{\Omega_s}\coh{23}+\Gamma\sqrt{d}\mathcal{E}\coh{41}+\left(\Delta-\Delta_s\right)\coh{43}\right)-\frac{\Gamma}{24}\left(\sqrt{3}\left(\coh{33}+\coh{44}\right)+10\coh{43}\right)
\end{split}\\
\begin{split}
\partial_z\mathcal{E}&=i\sqrt{d}\coh{13}
\end{split}
\end{align}
\end{subequations}

These equations were then numerically solved in XMDS2 with physical experimental parameters.  The full model plotted in figure \ref{fig:evp}a was obtained by sweeping the signal energy and extracting the phase by looking at the complex argument of the electric field, $\mathcal{E}$, using the following initial population and coherence parameters:

\begin{subequations}
\begin{align}
\begin{split}
\coh{11} (0)&=0.98
\end{split}\\
\begin{split}
\coh{22} (0)&=0.02
\end{split}\\
\begin{split}
\coh{33} (0)&=\coh{44}(0)=0
\end{split}\\
\begin{split}
\coh{12} (0)&=\coh{42}(0)=\coh{41}(0)=\coh{43}(0)=0
\end{split}
\end{align}
\end{subequations}

\section{Measured and theoretical signal absorption}

\begin{figure}[h]
\centering
\includegraphics[scale=0.9]{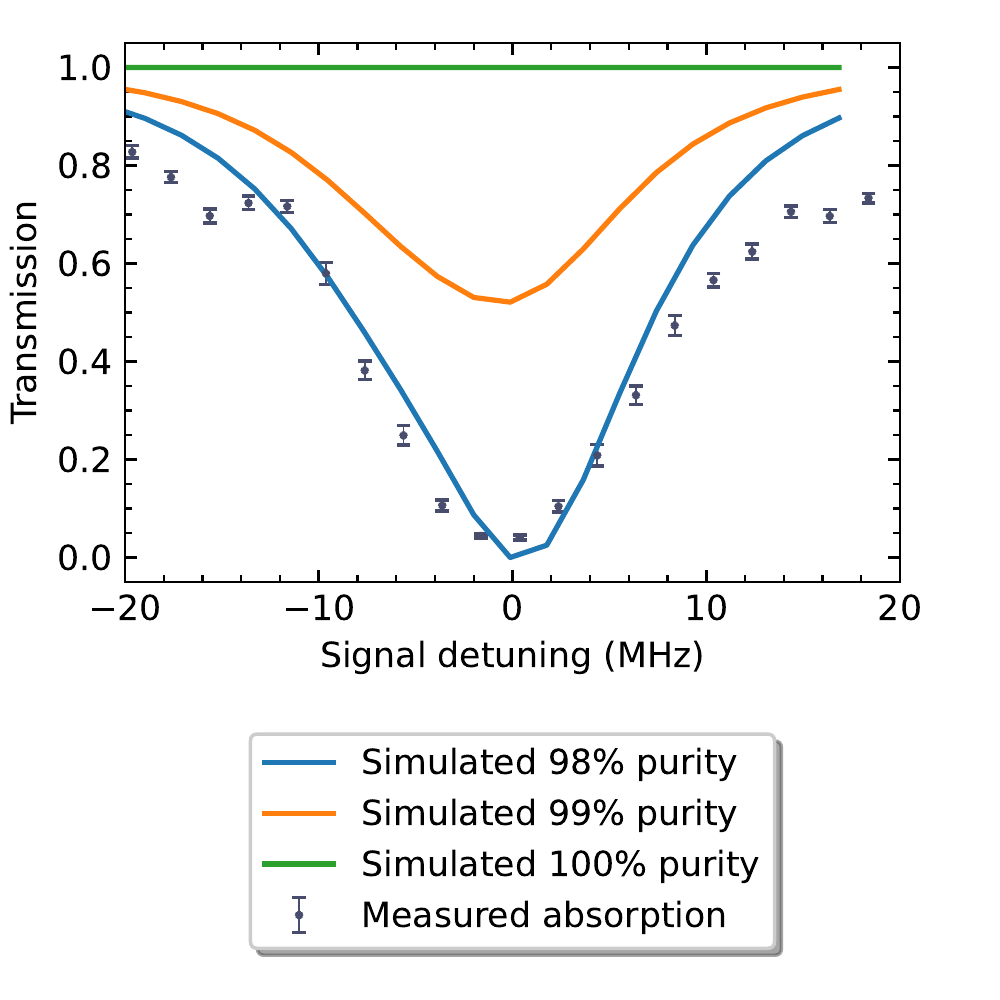}
\caption{Measured signal absorption along with theoretical signal absorption for 100\%, 99\% and 98\% initial state purity.}
\label{fig:sigabs}
\end{figure}


\end{document}